\documentclass[aps,preprint,preprintnumber,amsmath,amssymb,bm,12pt]{revtex4}
\usepackage[pdftex]{graphicx,color}
\usepackage{bm,amsmath,amssymb}
\usepackage{ascmac}

\begin{document}


{
\small
\rightline{
\baselineskip16pt\rm\vbox to20pt{
       \hbox{OCU-PHYS-483}
       \hbox{AP-GR-148}
}
}

\author{
Hideki~Ishihara$^1$}\email{ishihara@sci.osaka-cu.ac.jp}
\author{Ken~Matsuno$^{1,2}$}\email{matsuno@sci.osaka-cu.ac.jp}
\author{Masaaki~Takahashi$^{3}$}\email{mtakahas@auecc.aichi-edu.ac.jp} 
\author{Syuto~Teramae$^{1}$ \bigskip}\email{teramae@sci.osaka-cu.ac.jp}

\affiliation{${}^1$ 
Department of Mathematics and Physics, Graduate School of Science, Osaka City University,
Sumiyoshi, Osaka 558-8585, Japan
}
\affiliation{${}^2$ 
Faculty of Health Sciences, Butsuryo College of Osaka, Sakai, Osaka 593-8328, Japan 
}
\affiliation{~}
\affiliation{${}^3$ 
Department of Physics and Astronomy, Aichi University of Education, 
Kariya, Aichi 448-8542, Japan
\bigskip
}

\vskip 1cm

\title{
{\Large Particle acceleration by ion-acoustic solitons in plasma}\\ }

\begin{abstract}
We propose a new acceleration mechanism for charged particles by using cylindrical or spherical 
nonlinear acoustic waves propagating in ion-electron plasma. 
The acoustic wave, which is described by the cylindrical or spherical Kortweg-de Vries equation, 
grows in its wave height as the wave shrinks to the center. 
Charged particles confined by the electric potential accompanied with the shrinking wave get energy 
by repetition of reflections. 
We obtain power law spectrum of energy for accelerated particles. As an application, 
we discuss briefly that high energy particles coming from the Sun are produced 
by the present mechanism. 
\end{abstract}

\maketitle


\section{Introduction}

It is known that the energy spectrum of cosmic rays is well described 
by power laws over a very large energy span \cite{Gaisser}. 
It suggests a nonthermal acceleration mechanism of the high energy particles 
by a variety of active astrophysical objects: the solar atmosphere, supernova remnants, 
central region of galaxies, and so on. 
However, the acceleration mechanisms, which are important to understand the properties of 
the astrophysical objects, have not yet been elucidated. 

One of the most well-studied acceleration mechanisms is the Fermi acceleration \cite{Fermi:1949ee}, 
where charged particles gain nonthermal energy by repetition of reflections stochastically 
by magnetic clouds in astrophysical shock waves. 
By these multiple reflections the resulting energy spectrum of many particles becomes a power law. 
As alternative possibilities, a lot of mechanisms concerning to 
magnetic reconnections\rule[1mm]{4mm}{.5pt}double layer, monopole induction, 
and shock wave (surfing effect), etc.\rule[1mm]{4mm}{.5pt}are studied \cite{Aschwanden}. 
We consider a new acceleration mechanism by nonlinear acoustic solitonlike waves 
excited in a plasma. 

We consider a collisionless plasma of cold ions and isothermal electrons.  
It is well known that the one-dimensional planar ion-acoustic waves in the plasma are governed 
by the Kortweg-de Vries (KdV) equations \cite{Washimi-Taniuti}. 
In fact, such waves are really observed 
experimentally in the plasma system \cite{Ikezi-Taylor-Baker}. 
	
For cylindrical and spherical ion-acoustic waves in the plasma,  
modified KdV equations are introduced by  
Maxon and Viecelli \cite{Maxon-Viecelli1974c, Maxon-Viecelli1974s},  
and they showed existence of cylindrical and spherical solitonlike 
solutions by numerical calculations. 
While the planar solitons propagate with constant wave heights, 
the cylindrical and spherical solitons grow in their wave heights during the propagation toward the center.
Indeed, these waves are studied by numerical calculations of basic equations describing plasma 
systems \cite{Ogino-Takeda1976, Sheridan2017}, 
and also observed in laboratories 
\cite{Hershkowitz-Romesser, Ze-Hershkowitz-Chan-Lonngren}.

If the density fluctuation appears in the system of cold ions and warm electrons, 
the extent of electron density is broader than that of ion density. 
This means that positive charge excess occurs in the high density region.  
Therefore, an electric field is produced. 
The inhomogeneity of density accompanied with the electric field, described by the scalar potential field,  
propagates as an acoustic wave. 
Suppose that charged test particles (protons) are confined in the electric potential wall associated 
with the cylindrical or spherical ion-acoustic waves; 
the charged particles get energy after some reflections by moving the potential wall 
as the waves shrink into the center. An accelerated particle escapes 
from the potential wall as an output when the energy of the particle exceeds 
the electric potential energy.

In this paper, we present a new mechanism for the acceleration of charged particles 
by using nonlinear solitonlike acoustic waves propagating in plasma described 
by the cylindrical or spherical KdV equation. 
We show that the power law spectrum for accelerated output particles is obtained.
As an application, we briefly discuss a possibility that high-energy particles coming from 
the Sun are produced by the present acceleration mechanism. 

The organization of this paper is as follows. 
In the next section, we present the basic system considered in this study and derive the 
modified KdV equation that describes the cylindrically or spherically symmetric 
ion-acoustic waves. Then, we show the properties of solitonlike solutions to the equation.  
In Sec. \ref{acceleration}, we introduce a thin shell wall model to mimic the cylindrical or 
spherical soliton solution. Using the model, we trace a number of charged test particle motions 
accelerated by the soliton numerically  
and obtain the power spectrum of output particles. 
Section \ref{summary} is devoted to summary and discussion including an application to the solar cosmic rays. 

\section{Basic system}
\subsection{Basic equations of plasma}

We consider a plasma that consists of ions and electrons. 
The dynamics of the ions is described by a set of equations:
\begin{align}
	&M n^{(i)} 
		\left(\frac{\partial{\bm v}^{(i)}}{\partial t}
		+(\bm v^{(i)}\cdot \nabla) \bm v^{(i)} \right) 
		= en^{(i)}( {\bm E}+ {\bm v^{(i)}} \times {\bm B})-\nabla P^{(i)}, 
\label{eom_ion}
\\
	&\frac{\partial n^{(i)}}{\partial t} + \nabla \cdot (n^{(i)} {\bm v^{(i)}})=0 , 
\label{eoc_ion}
\end{align}
where $n^{(i)}, \bm v^{(i)}, P^{(i)}$ are  
the number density, the velocity, and the pressure of the ion fluid, and $M$ is the mass of ion. 
The electric and magnetic fields are denoted by $\bm E$ and $\bm B$, 
and $e$ is the elementary charge.  

We assume that there exists no global magnetic field, and neglect the magnetic field 
produced by the plasma motion\cite{Kuramitsu-Hada}. 
The electric field $\bm E$ is described by 
$\bm E= -\nabla \phi $, 
and the electric potential $\phi$ is governed by the Poisson equation, 
\begin{align}
	\triangle \phi = - \frac{e}{\varepsilon_0}(n^{(i)} - n^{(e)}) , 
\label{Poisson_eq}
\end{align}
where $n^{(e)}$ is the number density of electrons, and $\varepsilon_0$ is the 
vacuum permittivity. 

The electrons are assumed to be in thermal equilibrium with the temperature $T^{(e)}$, 
so that $n^{(e)}$ is given by 
\begin{align}
	n^{(e)} = n_0 ~{\rm exp}\left(\frac{e\phi}{k_B T^{(e)}}\right),
\end{align} 
where $k_B$ is the Boltzmann constant, and $n_0$ is 
the homogeneous density of electrons for $\phi=0$. 
Furthermore, we consider the case in which $P^{(i)}$ is negligible, 
namely the ions are cold.

\subsection{Cylindrical KdV and spherical KdV equations}

In this paper, we concentrate on nonlinear acoustic waves 
with cylindrical or spherical symmetry;  
then the basic equations \eqref{eom_ion} and \eqref{eoc_ion} are rewritten as
\begin{align}
	& \frac{\partial v^{(i)}}{\partial t}
		+v^{(i)} \frac{\partial v^{(i)}}{\partial r}  
		= - \frac{e}{M} \frac{\partial \phi}{\partial r}, 
\label{eom_ion_symm}
\\
	&\frac{\partial n^{(i)}}{\partial t} 
	+ \frac{\partial}{\partial r} (n^{(i)} v^{(i)})
	+\frac{2\gamma}{r} n^{(i)} v^{(i)}=0 , 
\label{eoc_symm}
\end{align}
and the Poisson equation \eqref{Poisson_eq} reduces to
\begin{align}
	\frac{\partial^2 \phi}{\partial r ^2}  
	+ \frac{2\gamma}{r}\frac{\partial \phi}{\partial r}  
	= - \frac{e}{\varepsilon_0}\left(n^{(i)} - n_0 ~{\rm exp}\left(\frac{e\phi}{k_B T^{(e)}}\right)\right), 
\label{Poisson_eq_symm}
\end{align}
where $\gamma=1/2$ for the cylindrical case, and $\gamma=1$ for the spherical case, respectively. 
The independent variable $r$ is the radial coordinate in the cylindrical or spherical 
coordinate system. 

For the purpose of taking acoustic waves that shrink toward the center into consideration, 
according to the reductive perturbation method \cite{Washimi-Taniuti}, 
we introduce new variables 
\begin{align}
	\xi &= \frac{\epsilon^{1/2}}{\lambda_D} (r + c_0 t), 
\label{xi}
\\
	\tau &=\frac{\epsilon^{3/2}}{\lambda_D} c_0 t ,
\label{tau}
\end{align}
where $\epsilon$ is a small constant, $\lambda_D $ is the Debye length given by
\begin{align}
	\lambda_D = \sqrt{\frac{\varepsilon_0 k_B T^{(e)} }{n_0 e^2}}, 
\end{align}
and $c_0$ is the sound velocity defined by  
\begin{align}
	c_0=\sqrt{\frac{k_B T^{(e)}}{M}}. 
\label{c_0}
\end{align}

We rewrite Eqs.\eqref{eom_ion_symm}-\eqref{Poisson_eq_symm} with these variables as
\begin{align}
	& \epsilon^{1/2}c_0 \frac{\partial v^{(i)}}{\partial\xi} + \epsilon^{3/2}c_0 \frac{\partial v^{(i)}}{\partial\tau} 
	+ \epsilon^{1/2} v^{(i)} \frac{\partial v^{(i)}}{\partial\xi}
		= - \epsilon^{1/2} \frac{e}{M} \frac{\partial \phi}{\partial\xi} , 
\label{eom_ion_xitau}
\\
	& \epsilon^{1/2} c_0\frac{\partial n^{(i)}}{\partial\xi} 
	+\epsilon^{3/2}c_0 \frac{\partial n^{(i)}}{\partial\tau}
	+ \epsilon^{1/2} \frac{\partial}{\partial\xi}  (n^{(i)} v^{(i)})
	- \epsilon^{3/2} \frac{2\gamma}{(\tau-\epsilon \xi)} n^{(i)} v^{(i)}=0 , 
\label{eoc__xitau}
\\
	& \epsilon \frac{\partial^2 \phi}{\partial\xi^2}   
	- \epsilon^2 \frac{2\gamma}{(\tau-\epsilon \xi)} \frac{\partial \phi}{\partial\xi}  
	= - \lambda_D^2\frac{e}{\varepsilon_0}\left(n^{(i)} 
		- n_0 ~{\rm exp}\left(\frac{e\phi}{k_B T^{(e)}}\right)\right).
\label{Poisson_eq_xitau}
\end{align}
We expand variables $v^{(i)}, \phi$ and $n^{(i)}$ by $\epsilon$ in the form
\begin{align}
	\frac{v^{(i)}}{c_0} =& \epsilon v_1 +\epsilon ^2 v_2 + \cdots, 
\label{exp_v}
\\
	\frac{e \phi}{k_B T^{(e)}}=&\epsilon \phi_1 +\epsilon ^2 \phi_2+ \cdots, 
\label{exp_phi}
\\
	\frac{n^{(i)}}{n_0} =& 1 + \epsilon n_{1} + \epsilon^2 n_2 + \cdots .
\label{exp_n}
\end{align}
Substituting Eqs.\eqref{exp_v}-\eqref{exp_n}
into Eqs.\eqref{eom_ion_xitau}-\eqref{Poisson_eq_xitau}, 
we obtain a set of equations order by order in $\epsilon$. 
The lowest equations in $\epsilon$ are
\begin{align}
 	n_1=-v_1=\phi _1, 
\label{lowest}
\end{align}
and the second lowest order equations give  
\begin{align}
  	\frac{\partial n_1}{\partial \tau}
	-\frac{\partial v_1}{\partial \tau}
	-v_1\frac{\partial v_1}{\partial \xi}
	+\phi_1\frac{\partial \phi_1}{\partial \xi}
	+\frac{\partial }{\partial \xi}(n_1 v_1)
	- \frac{\partial^3 \phi_1}{\partial \xi ^3}
	 - 2 \gamma \frac{v_1}{\tau}=0 .
\label{second_eq}
\end{align}
From Eqs.\eqref{lowest} and \eqref{second_eq} we obtain
\begin{align}
  \frac{\partial \Phi}{\partial \tau}
	-\Phi\frac{\partial \Phi}{\partial \xi} 
	-\frac{1}{2} \frac{\partial ^3 \Phi}{\partial \xi ^3}
	+ \gamma \frac{\Phi}{\tau}=0,
\label{ext_KdV}
\end{align}
where $\Phi:= \phi_1=-v_1=n_1$.
If $\gamma=0$, Eq.\eqref{ext_KdV} is the KdV equation, which describes nonlinear 
plane waves. In the case $\gamma = 1/2$ or $1$ the equation is the extended 
KdV equation that describes cylindrical or spherical waves, respectively. 
From Eqs.\eqref{xi} and \eqref{tau} we see
\begin{align}
	r =\lambda_D(\epsilon^{-1/2}\xi- \epsilon^{-3/2}\tau) ,
\end{align}
then $r=0$ corresponds to $\tau=0$ in the lowest order with respect to $\epsilon$. 
The cylindrical or spherical wave shrinks from an initial radius $r=r_0$ to $r=0$ 
as increasing $\tau$ from the initial time $\tau=\tau_0<0$ to $\tau=0$. 

\subsection{ Properties of cylindrical and spherical soliton solutions}

We study characteristic properties of solitonlike wave solutions 
with cylindrical or spherical symmetry. 
In the case of $\gamma=0$, 
it is well known that the KdV equation has soliton solutions in the form
\begin{align}
	\Phi = A~{\rm sech}^2\left[\sqrt\frac{A}{6} 
		\left(\xi +\frac{A}{3} \tau\right)\right] , 
\label{planar_soliton}
\end{align}
where the wave height denoted by $A$ is a constant. The soliton described by the solution 
\eqref{planar_soliton} propagates 
with the constant velocity $A/3$ in the $\xi$-$\tau$ plane, keeping its shape invariant. 

In the cylindrical or spherical case, $\gamma=1/2$ or $1$,  
we set a wave with the radius $r=r_0$ and the width is much smaller than $r_0$ 
at the initial time.
In this set up, the cylindrical or spherical wave is 
described approximately by the planar wave Eq.\eqref{planar_soliton}. 
However, the wave height is no longer constant owing to the existence of the last term 
in Eq.\eqref{ext_KdV}. As is shown later, the wave height grows in time as the wave shrinks 
toward the center. 
Numerical solutions to the cylindrical KdV and the spherical KdV equations are widely studied~ 
\cite{Maxon-Viecelli1974c, Maxon-Viecelli1974s, 
Ko-Kuehl1979,Hase-Watanabe-Tanaca1985,Infeld-Rowlands2000} 
and showed the growth of the wave height.

For a wave on a finite support, Eq.\eqref{ext_KdV} admits a conserved quantity $Q$ in the form 
\begin{align}
	Q = |\tau|^{2\gamma}\int_{-\infty}^\infty \Phi^2 ~ d \xi .
\label{Q2}
\end{align}
After replacing the constant $A$ in Eq.\eqref{planar_soliton} by a function $A(\tau)$ we substitute it 
into Eq.\eqref{Q2},  
then we see the peak height of the waves grows as 
$(\tau/\tau_0)^{-4\gamma/3}$ while the width shrinks as $(\tau/\tau_0)^{2\gamma/3}$~
\cite{Ko-Kuehl1979}.

In the final stage, $\tau \sim 0$, of the cylindrical case, $\gamma=1/2$,  
we find that the time derivative term and the time dependent term, 
the first and the last terms,  
dominate the nonlinear term and the dispersive term, the second and third terms, 
in Eq.\eqref{ext_KdV} for a wide range of initial conditions of numerical calculations.  
Namely, Eq.\eqref{ext_KdV} becomes
\begin{align}
  \frac{\partial \Phi}{\partial \tau} + \frac12 \frac{\Phi}{\tau} \approx 0 , 
\label{final_KdV}
\end{align}
then we see that the wave height grows as
$\sim (\tau/\tau_0)^{-1/2}$ with a constant width.
On the other hand, in the final stage of the spherical case, $\gamma=1$, for numbers of initial conditions, 
we observe numerically that 
the contribution of the dispersive term becomes small, 
and the wave height grows as $\sim (\tau/\tau_0)^{-1}$.  
Figure \ref{fig:cylindrical_solitons} and Figure \ref{fig:spherical_solitons} show the numerical evolution 
of the wave forms of cylindrical and spherical solitons. 
Figure \ref{fig:t_dep_wave_height} shows examples of the time dependence of the wave height in both cases.

The amplitude $\Phi=\phi_1$ of the wave describes the electric potential produced by 
the charge excess at the peak of the wave. 
The cylindrical or spherical wave is accompanied by the cylindrical or spherical 
electric potential wall. 
We consider test charged particles that are confined by the potential wall. 
The moving charged particles are reflected by the shrinking potential wall, 
and the particles are accelerated. 
When the particle gets kinetic energy greater than the electric potential, 
the particle escapes from the region enclosed by the 
potential wall. 
The wave height of the cylindrical or spherical wave grows as the wall shrinks toward 
the center, the energy spectrum of escaped particles depends on the time evolution of the 
wave height.  

\begin{figure}[!htbp]
\begin{center}
\includegraphics[height=5cm]{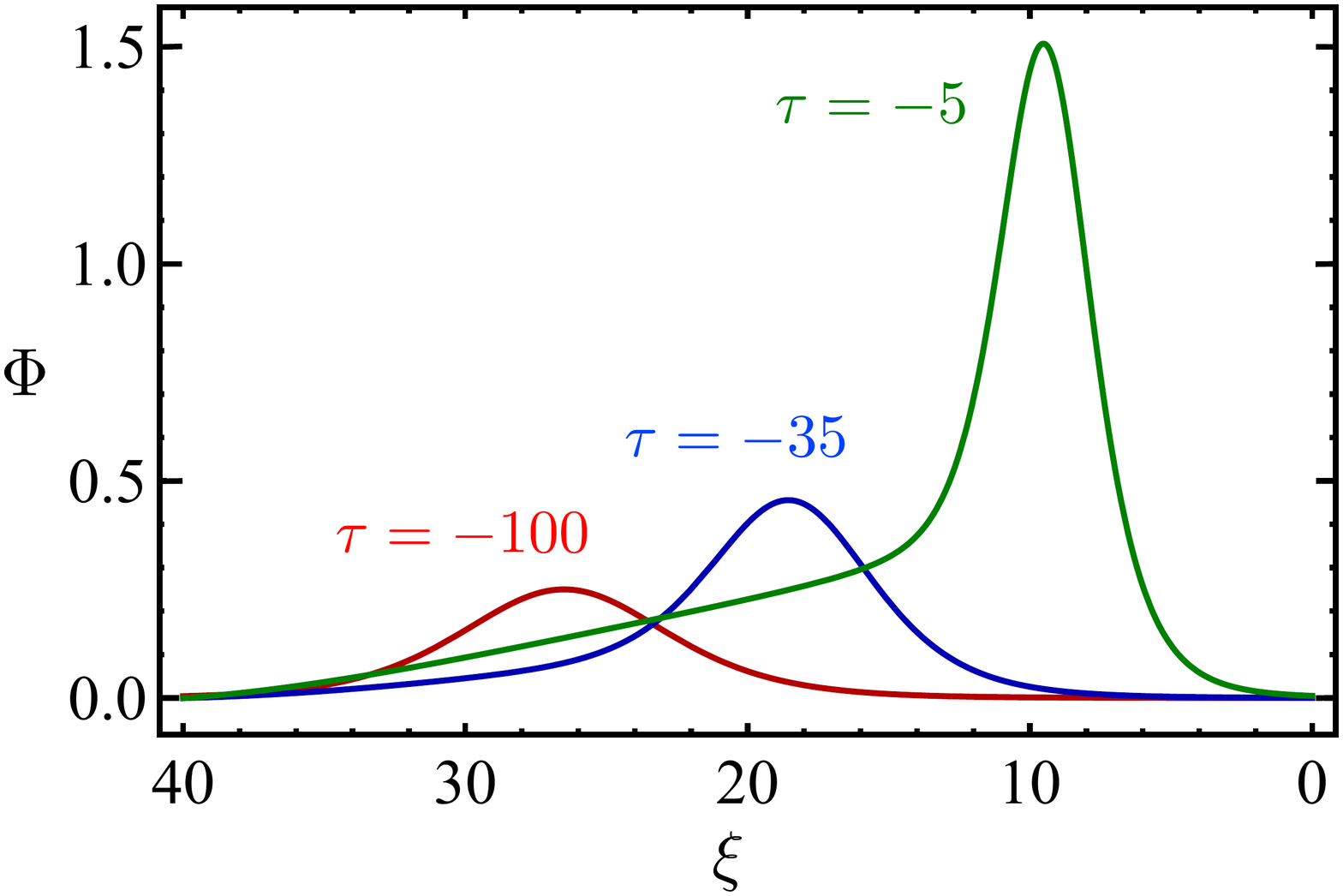}\qquad\qquad
\includegraphics[height=5.5cm]{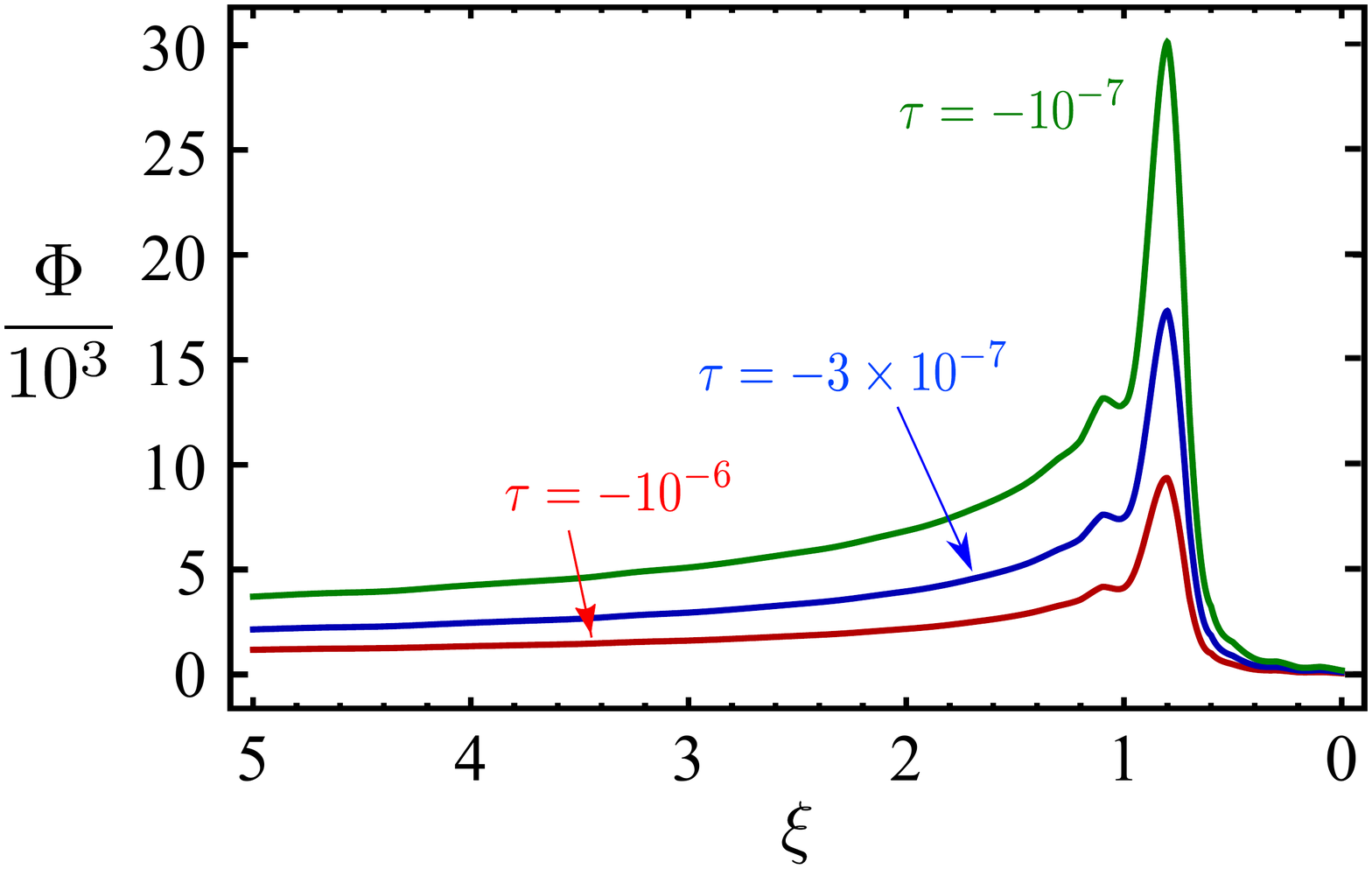}
\caption{Evolution of the cylindrical soliton. 
Wave forms in the early stage: $\tau=-100, -35, -5$ (left panel). 
Wave forms in the final stage: $\tau= -10^{-6}, 
-3\times 10^{-7}, -10^{-7}$ (right panel).
}
\label{fig:cylindrical_solitons}
\end{center}
\end{figure}

\begin{figure}[!htbp]
\begin{center}
\includegraphics[height=5cm]{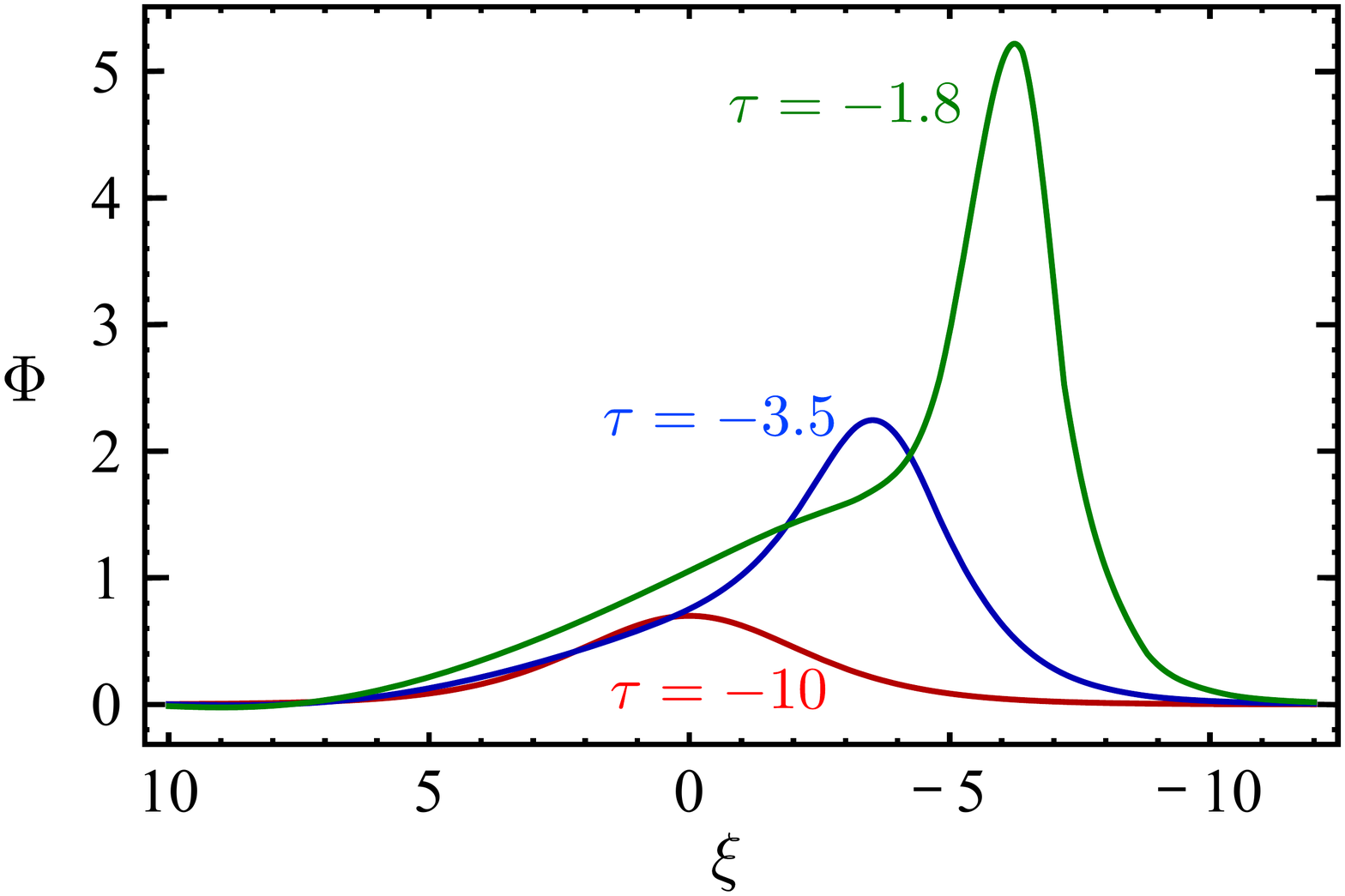}\qquad\qquad
\includegraphics[height=4.4cm]{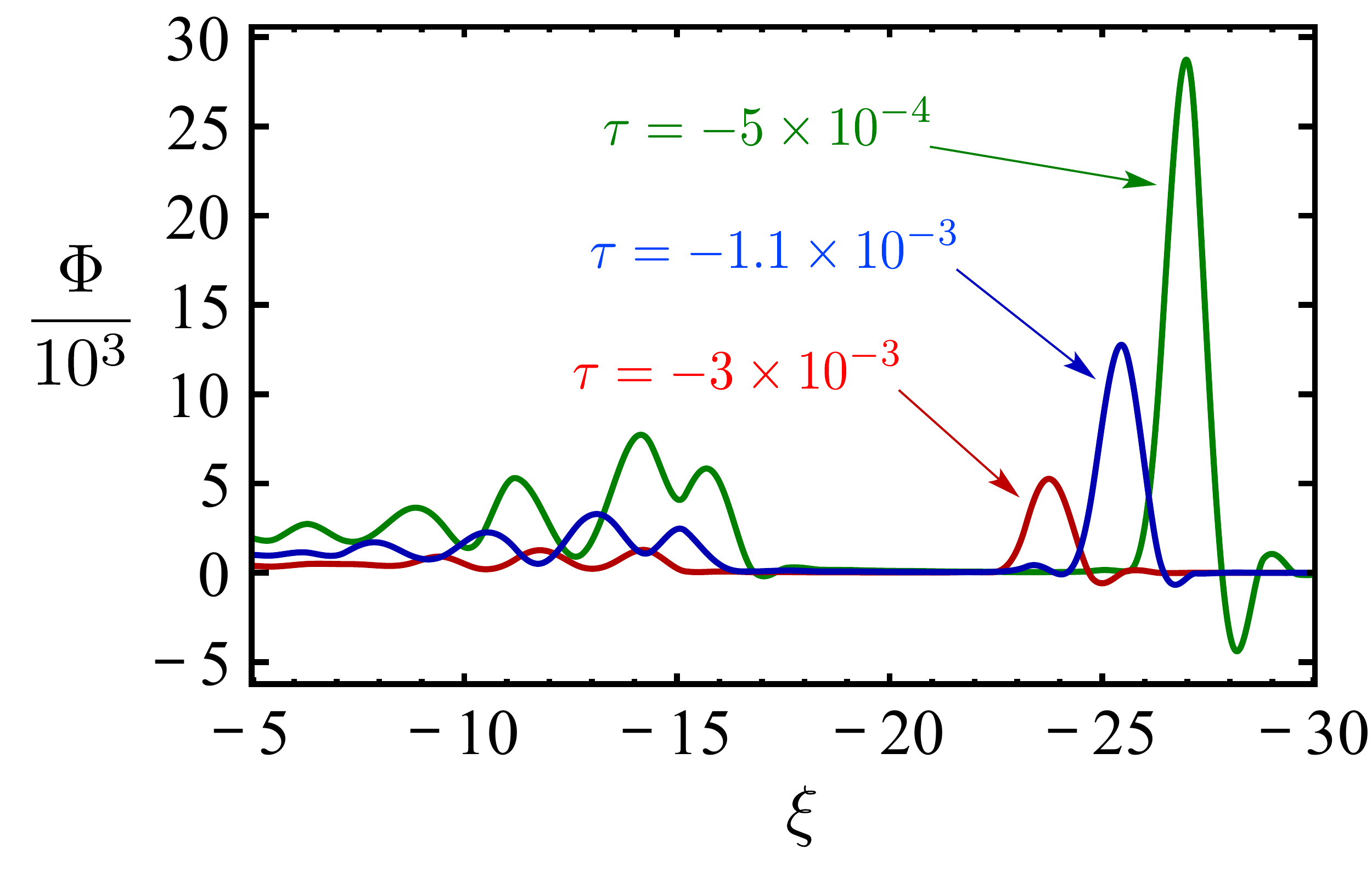}
\caption{Evolution of the spherical soliton.
Wave forms in the early stage: $\tau=-10, -3.5, -1.8$ (left panel). 
Wave forms in the final stage: 
$\tau=-3\times 10^{-3}, -1.1\times 10^{-3}, -5\times 10^{-4}$ (right panel).
}
\label{fig:spherical_solitons}
\end{center}
\end{figure}

\begin{figure}[!htbp]
\begin{center}
\includegraphics[height=5.5cm]{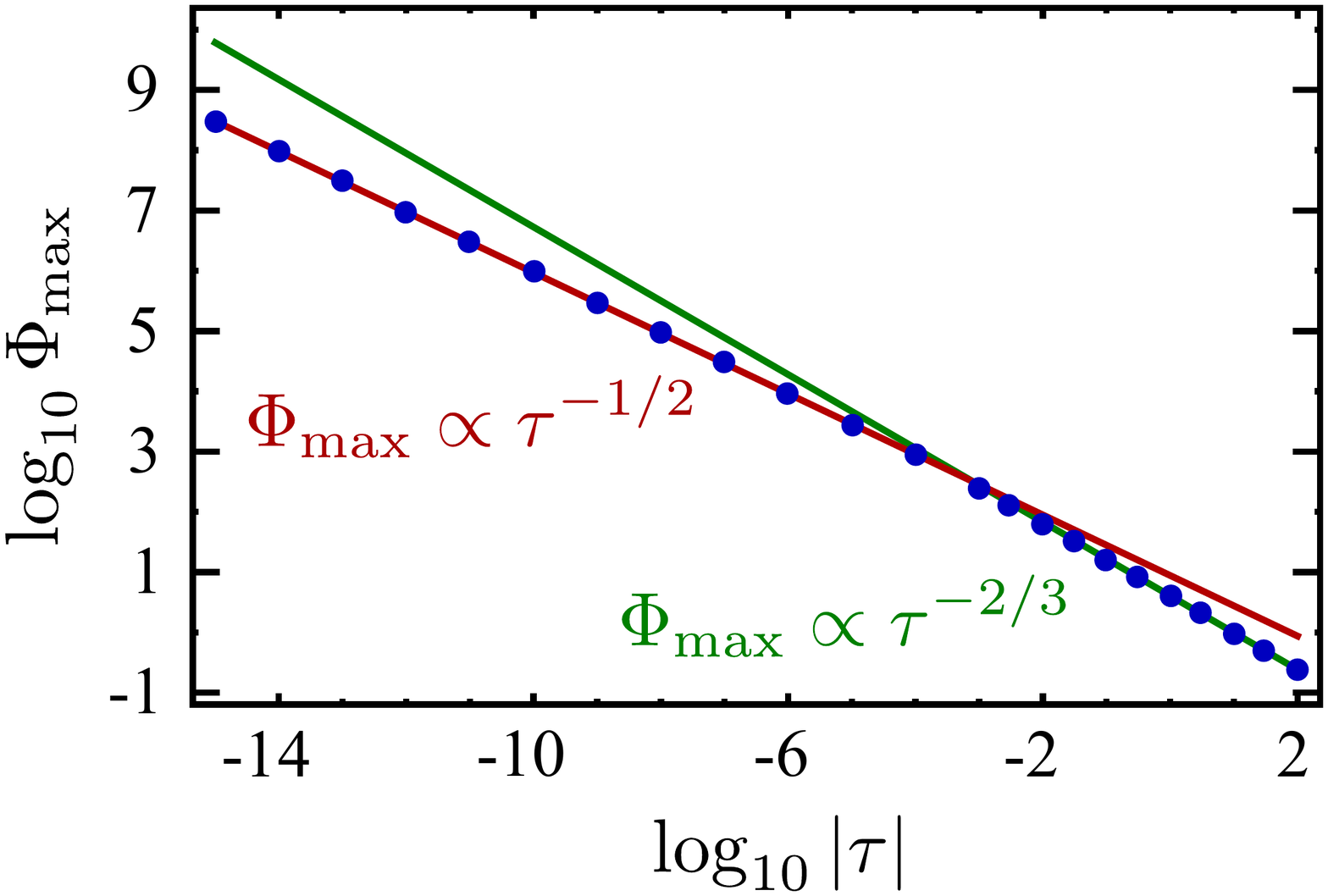} \qquad
\includegraphics[height=5.3cm]{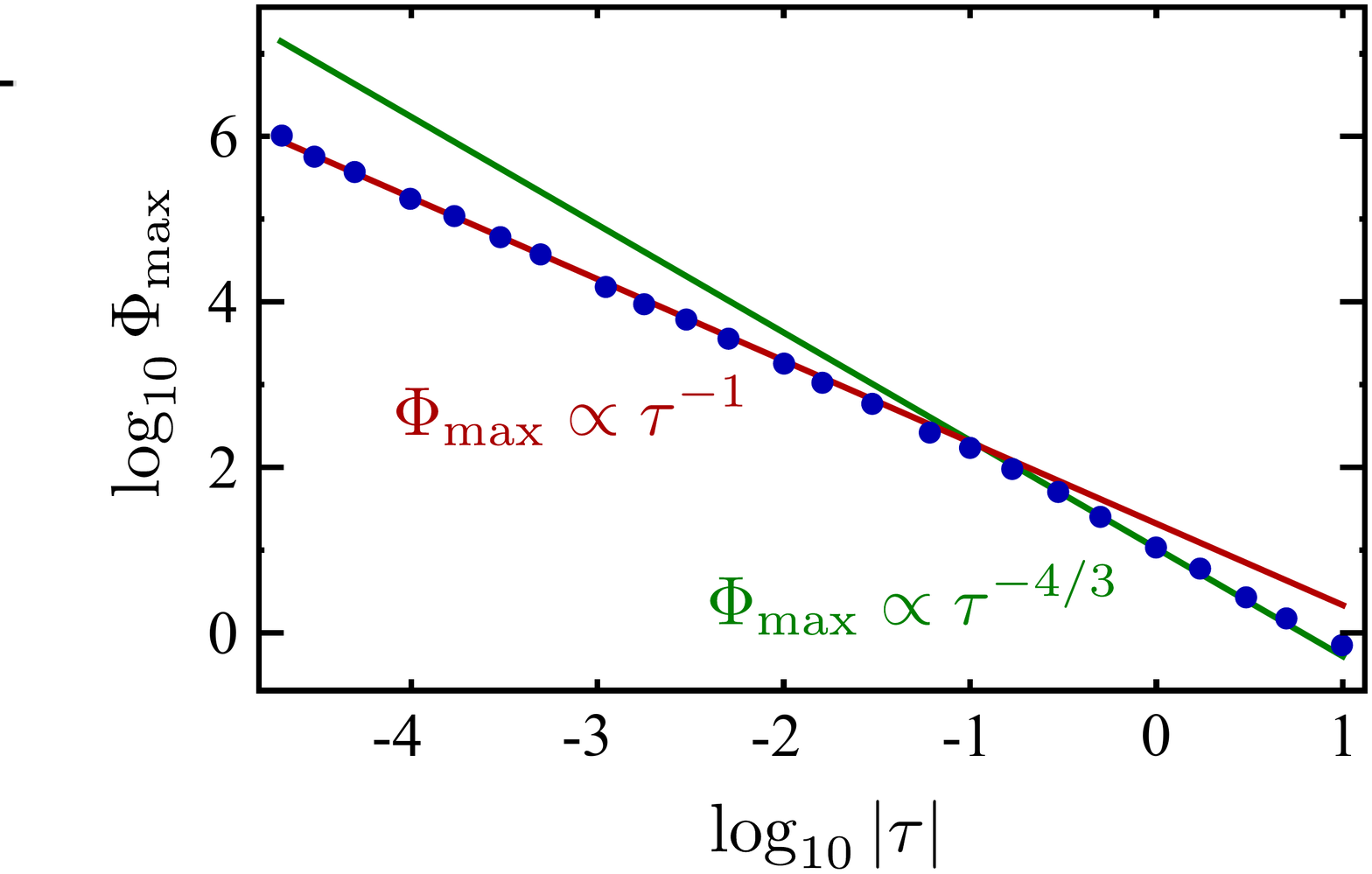}
\caption{Time evolution of the wave height $\Phi_{\rm max}$ for the cylindrical soliton (left panel). 
In the early stage $\Phi_{\rm max}\propto \tau^{-2/3}$, 
while in the final stage $\Phi_{\rm max}\propto \tau^{-1/2}$. 
The same one for the spherical soliton (right panel).  
In the early stage $\Phi_{\rm max}\propto \tau^{-4/3}$, 
while in the final stage $\Phi_{\rm max}\propto \tau^{-1}$. 
}
\label{fig:t_dep_wave_height}
\end{center}
\end{figure}

\section{Acceleration of particles}
\label{acceleration}

We consider that test charged particles are accelerated by the shrinking 
potential wall described by the cylindrical or spherical solitons. 
In order to simplify the system, we make a model that the soliton is 
replaced by a thin shell wall. We calculate test particle motion enclosed by 
this shrinking thin shell wall numerically and obtain the energy spectrum of 
the accelerated particles.

\subsection{Thin shell wall models}
In contrast to the plane soliton solution to the KdV equation, 
the most important property of the cylindrical or 
spherical soliton is that the wave height grows in time, $t$, as the wave goes to the center. 
We reduce the cylindrical or spherical soliton to a thin shell wall at the peak position of the wave,  
where the width of the wave is ignored. 
Furthermore, we ignore, here, the motion of the wave in the $\xi$-$\tau$ plane.
It means that the wave propagates with the speed $c_0$ in the $r$-$t$ plane. 
The thin shell wall describes the electric potential wall 
whose height evolves in time.

The model of the thin shell wall is specified by the following properties:
\begin{enumerate}
\item
The initial radius of the shell is $r_0$ at the initial time $t_0 (<0)$.  
We assume the speed of thin shell in the $r$-$t$ plane is the sound speed $c_0$, 
then the radius of the shell is described by $r(t) = - c_0 t$. 
\item
According to the growth rate of the wave height of the cylindrical or spherical soliton 
discussed in the previous section, 
we assume that the height of the thin shell wall grows as $\Phi(t)=\Phi_0~(t/t_0)^{-\alpha}$, 
 $\alpha= 1/2$ or $2/3$ for the cylindrical case, and $\alpha= 1$ or $4/3$ for the spherical case, 
where $\Phi_0$ is the initial amplitude of electric potential.
\item
We should stop the thin shell wall evolution when the shell radius becomes the Debye length. 
Then the final time is given by $t_f= - \lambda_D/c_0$. 
\end{enumerate}

Motion of test charged particles is assumed as follows:
\begin{enumerate}
\item{\sl Elastic reflection}
\newline
A moving charged particle toward the thin shell wall with the velocity $\bm v=(v_\perp,~ v_\parallel)$
gets the velocity $\bm v=(-v_\perp-2c_0,~ v_\parallel)$ after a reflection by the shrinking wall 
with the sound velocity $c_0$, where $v_\perp$ and $v_\parallel$ are the velocity components of  
the normal and tangential to the thin shell wall, respectively. 
\item{\sl Collisionless}
\newline
We assume that each charged test particle moves with a constant velocity till it hits the 
thin shell wall, 
and the test particles do not collide with each other.

\item{\sl Particle escaping criterion} 
\newline
If the kinetic energy of a particle exceeds the height of the thin shell wall $\Phi(t)$, 
the wall cannot confine the particle then the particle escapes to the infinity 
as an output particle. 
\end{enumerate}

A typical trajectory of a test particle reflected by the shrinking thin shell wall 
is shown in Fig.\ref{fig:reflection}.

\begin{figure}[!htbp]
\begin{center}
\includegraphics[height=4.5cm]{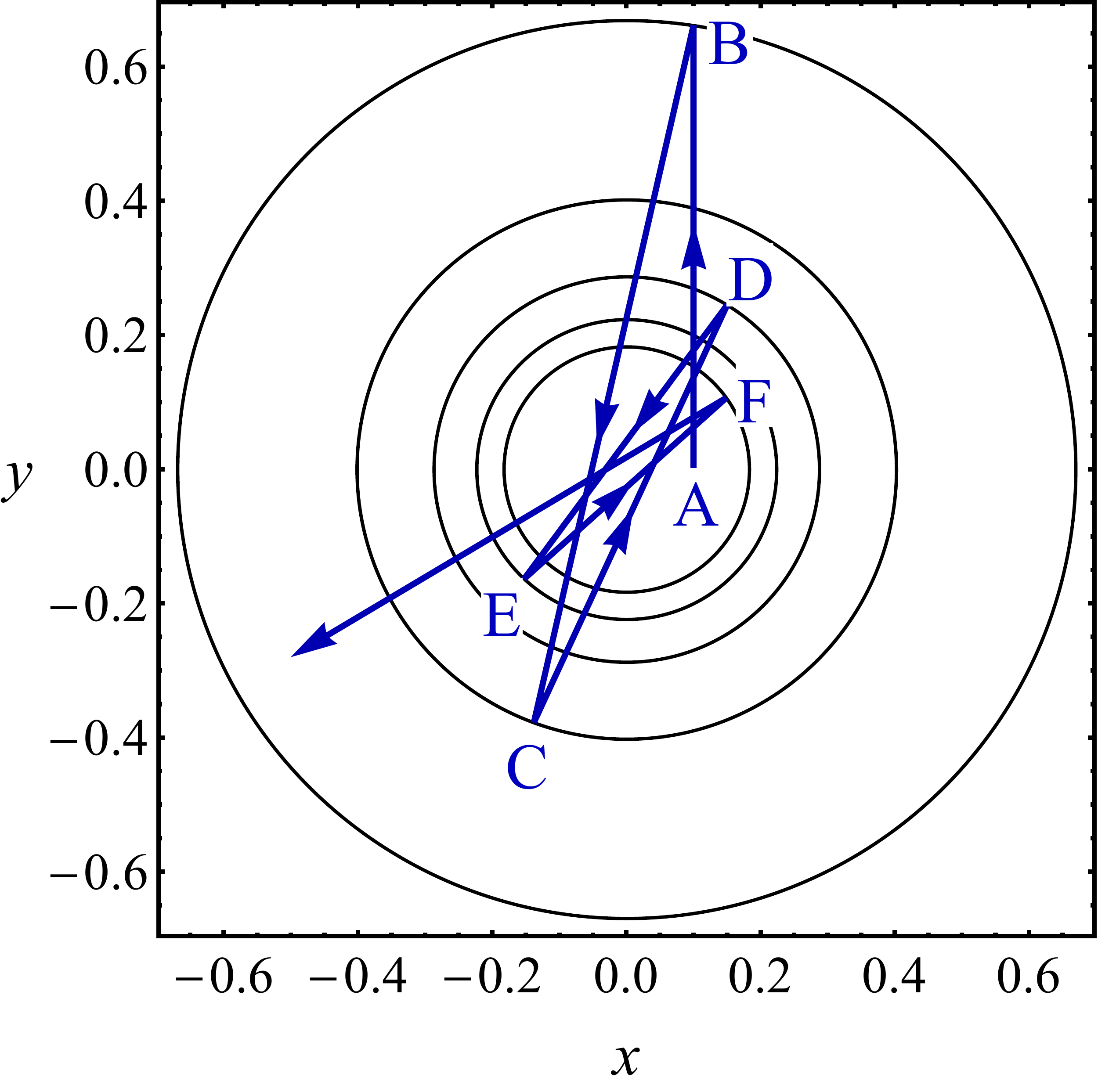}
\hspace{2cm}
\includegraphics[height=4cm]{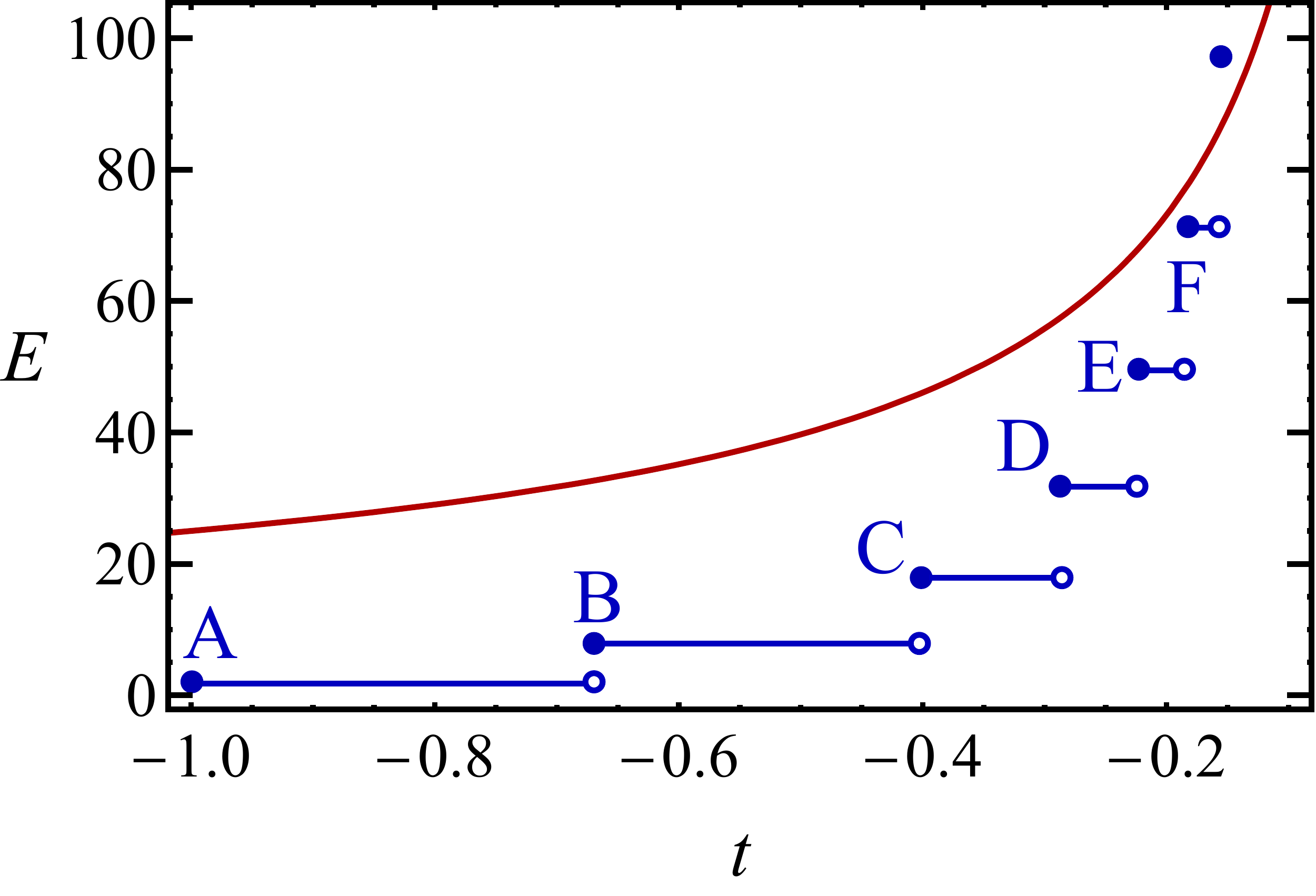}
\caption{A typical trajectory of a test particle in the cylindrical case (left panel). 
The particle is reflected elastically by the shrinking thin shell. 
Time evolution of electric potential, $\Phi\propto t^{-2/3}$, is drawn by 
the solid (red) curve, and particle energy gained by reflections is shown by bars (right panel). 
}
\label{fig:reflection}
\end{center}
\end{figure}

\subsection{Numerical studies for acceleration of particles}

We consider protons as ions, i.e., $M$ is the proton mass $M_P$, 
and settle a thin shell wall initially with 
$r_0= 10^6 \lambda_D$ and  $\Phi_0=k_B T^{(e)}$.
The initial time and final time are given by 
$t_0= - r_0/c_0$ and $t_f=-\lambda_D /c_0$, where the sound velocity $c_0$ is 
given by Eq.\eqref{c_0}. 

Here, we consider the initial distribution of the test charged particles. 
We assume the Maxwell distribution with the temperature $T \leq T^{(e)}$ 
of test charged particles with a constant spatial density 
enclosed by the thin shell wall (see Fig.\ref{fig:Maxwell_dist}).  

\begin{figure}[!htbp]
\begin{center}
\includegraphics[height=4.3cm]{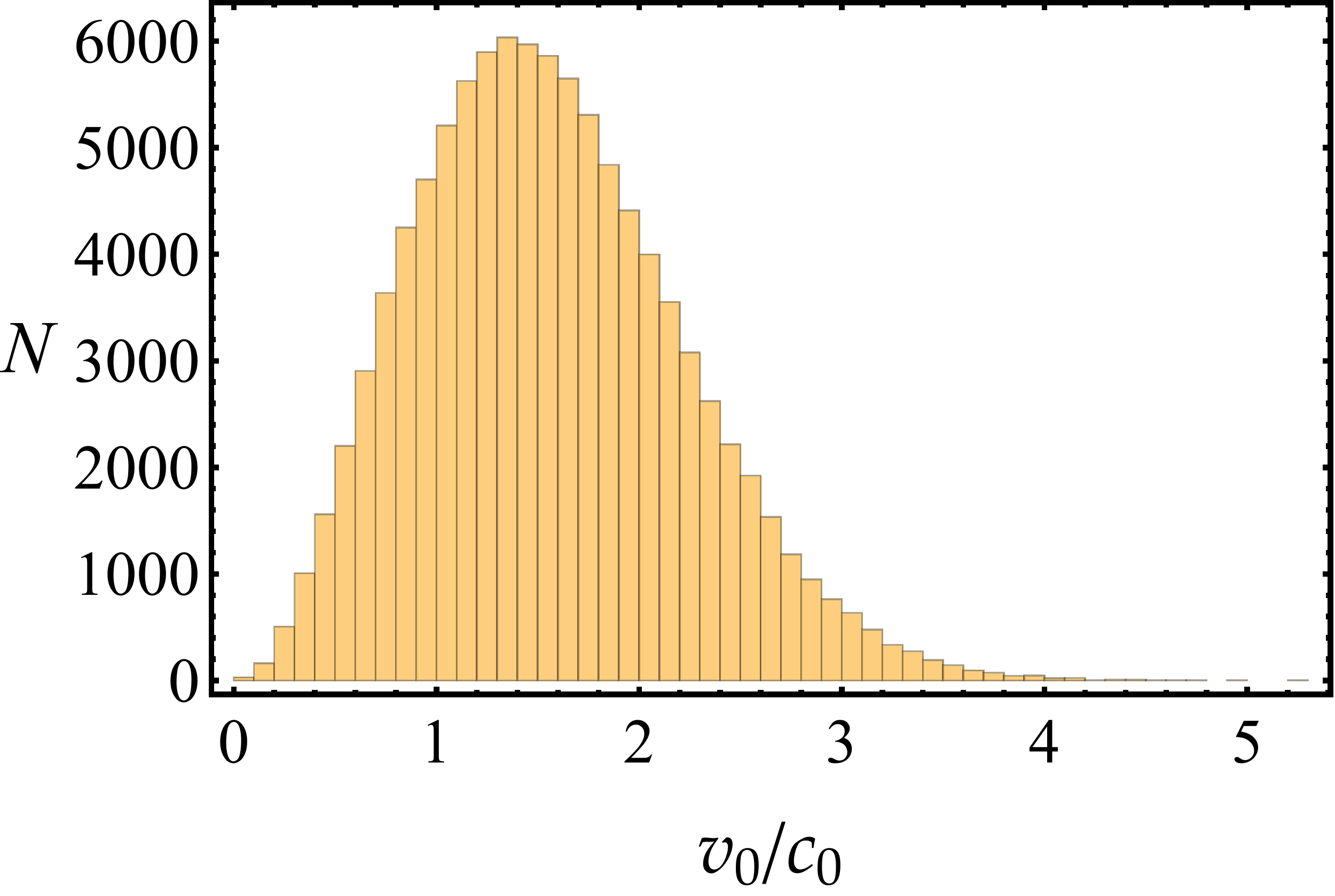}
\caption{Initial distribution function of $10^5$ test charged particles is 
assumed to be the Maxwell distribution. 
}
\label{fig:Maxwell_dist}
\end{center}
\end{figure}

We trace a number of test particles moving and reflected by the thin shell wall, 
and obtain the energy spectrum of output particles in 
$\alpha =2/3$ and $\alpha =1/2$ in the cylindrical case, and 
$\alpha =4/3$ and $\alpha =1$ in the spherical case. 
Figure \ref{fig:spectrum_cylindrical} and Figure \ref{fig:spectrum_spherical} show 
that the high-energy part of the energy spectrum is described 
by a power law, $E^{-p}$, in these models. 
Table~\ref{table:cylindrical} shows that the values of power index $-p$ for different $\alpha$ and several 
initial temperatures of the test particles $T$. 
In both cylindrical and spherical models, 
the power index $-p$ depends on $\alpha$, which determines the evolution of the electric potential height. 
However, it does not depend on the temperature of initial distribution of the test charged particles.

In the numerical experiments with $10^5$ initial test particles, 
the maximum energy of the output particle is $3.8\times 10^2 ~k_B T^{(e)}$ for $\alpha=2/3$, 
and $5.1\times 10 ~k_B T^{(e)}$ for $\alpha=1/2$ in the cylindrical model. 
The same one is $4.2 \times 10^6 ~k_B T^{(e)}$ for $\alpha=4/3$, and $2.1 \times 10^3 ~k_B T^{(e)}$ for $\alpha=1$ 
in the spherical model. 
All particles escape from the thin shell wall before $t=t_f$ in the present calculations. 
The power law spectrum of the output particles, 
which does not depend on the initial numbers of particles, has no characteristic scale of energy;  
then if we set much numbers of test particles 
initially, we can get more energetic output particles. 
The maximum energy is limited by the applicability of the soliton model. 
Particles can be accelerated till 
the radius of cylindrical or spherical solitons become the Debye length. 
Therefore, if the initial numbers of particles is large enough, we 
would obtain the energy $E_{max}=\Phi(t_f)$ as the maximum.

\begin{figure}[!h]
\begin{center}
\includegraphics[height=5cm]{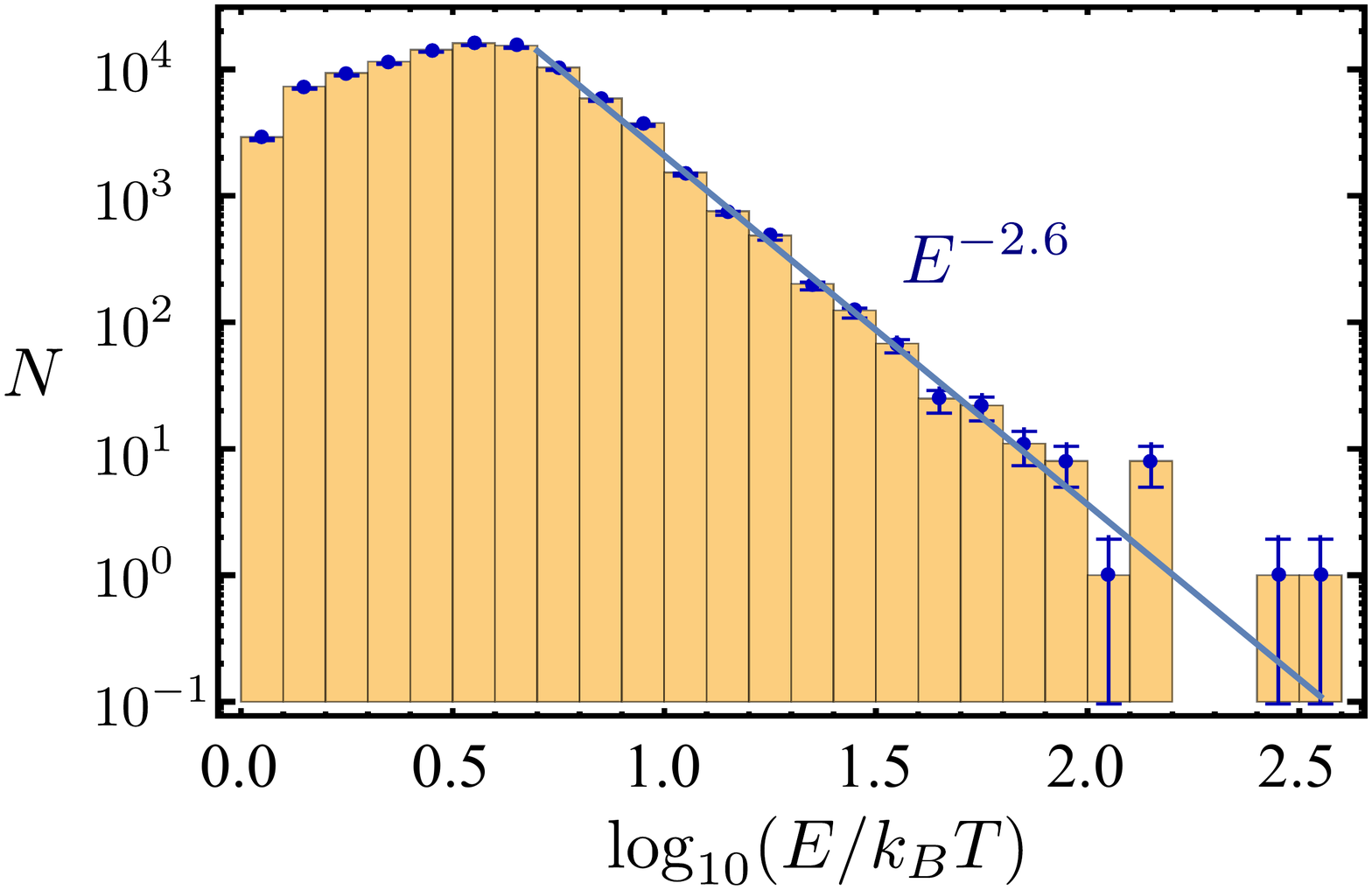}\qquad
\includegraphics[height=5cm]{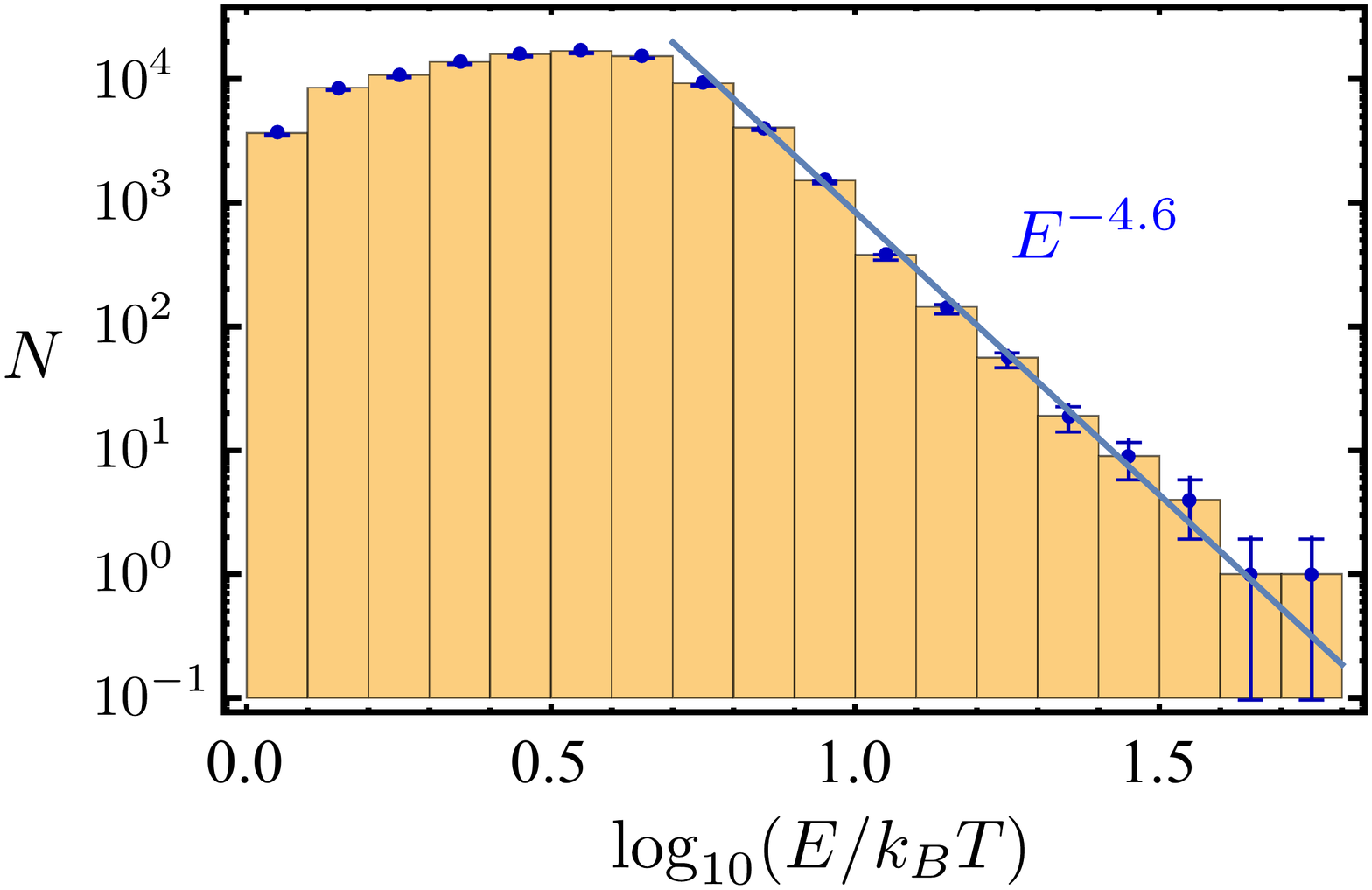}
\caption{Energy spectrum of output particles in the cylindrical model. 
The case of  $\alpha =2/3$ (left panel), and 
the case of  $\alpha =1/2$ (right panel). }
\label{fig:spectrum_cylindrical}
\end{center}
\end{figure}
\begin{figure}[!h]
\begin{center}
\includegraphics[height=5cm]{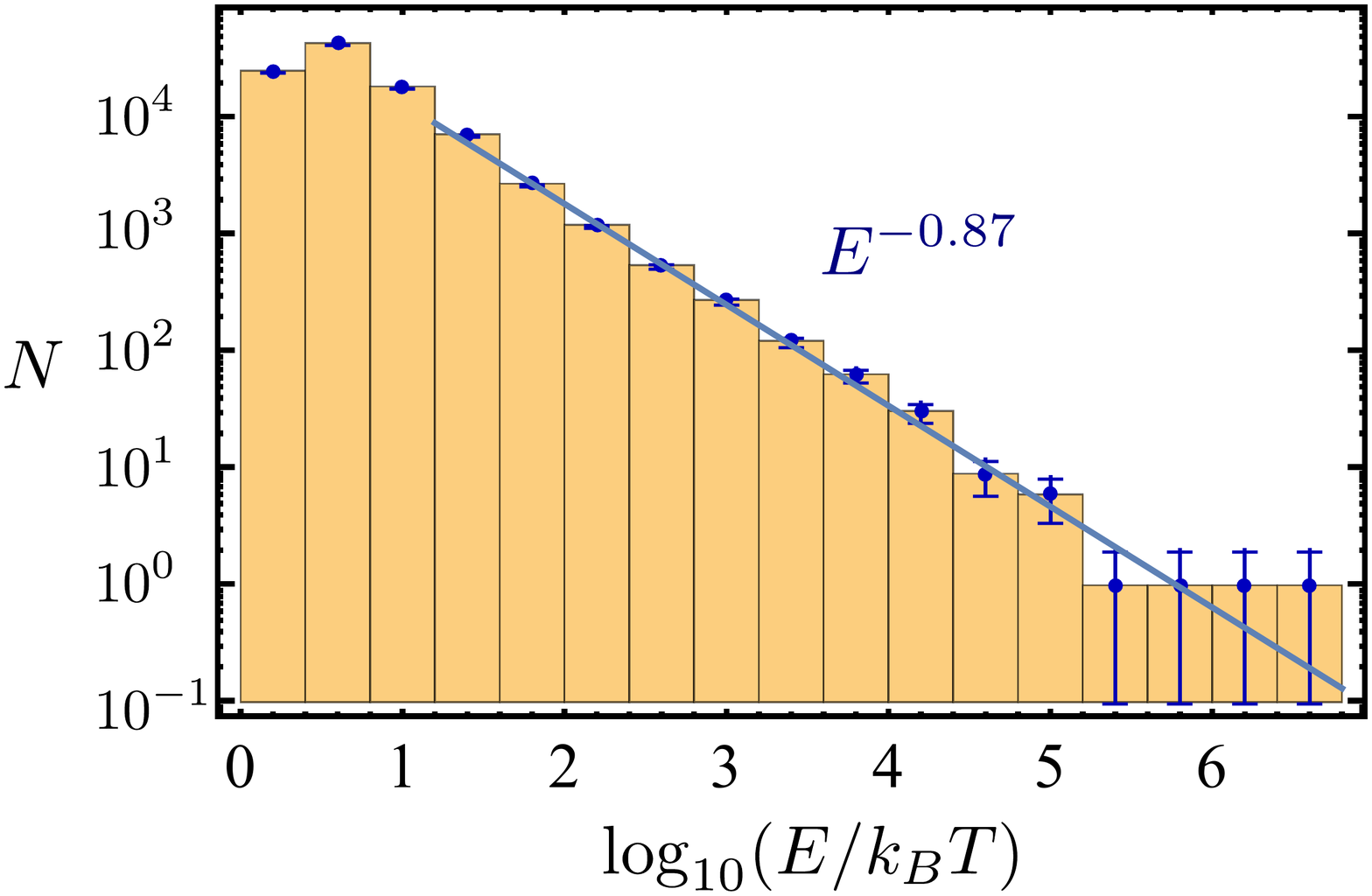}\qquad
\includegraphics[height=5cm]{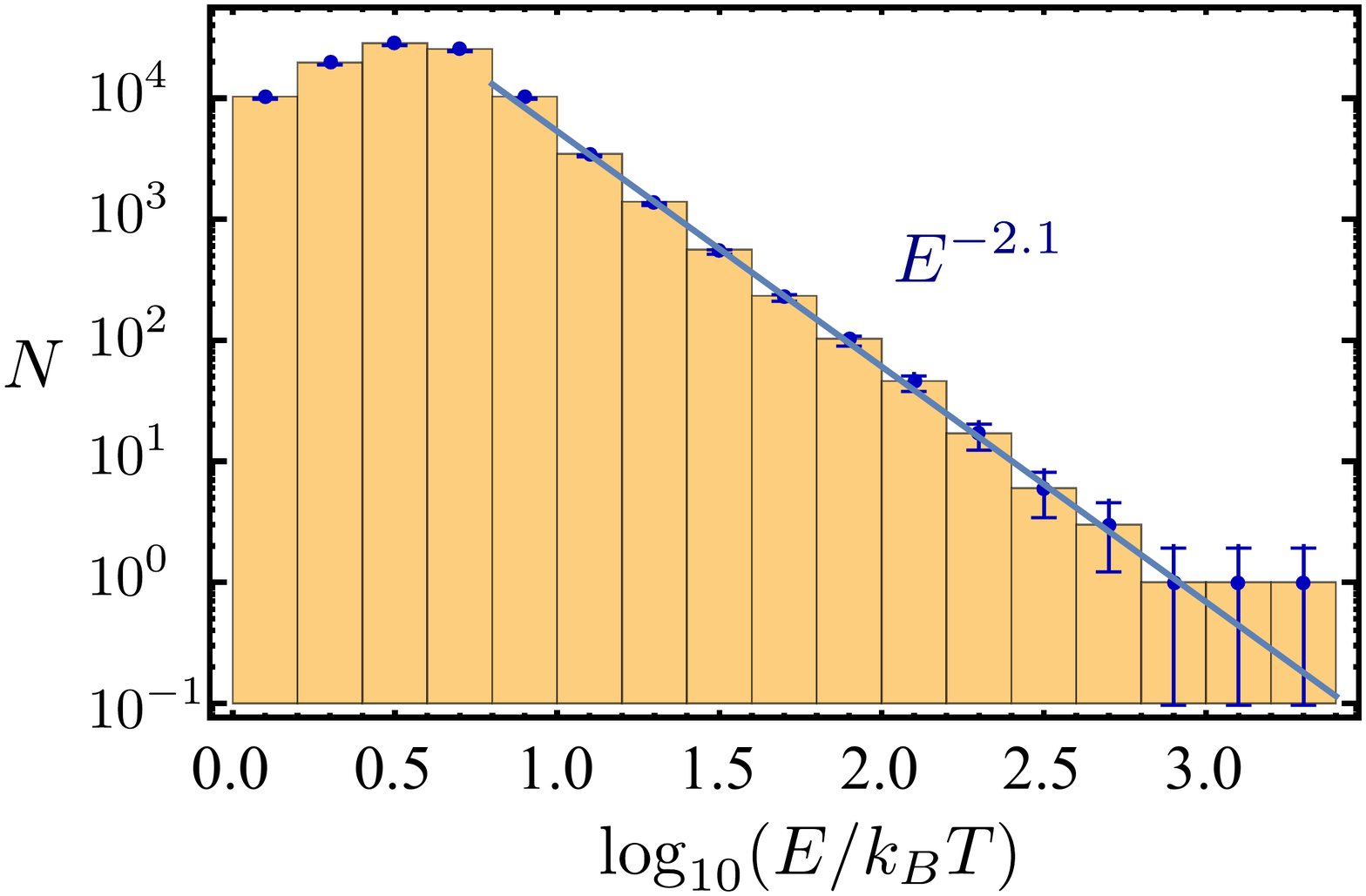}
\caption{Energy spectrum of output particles in the spherical model. 
The case of  $\alpha =4/3$ (left panel), and 
the case of  $\alpha =1$ (right panel). 
}
\label{fig:spectrum_spherical}
\end{center}
\end{figure}

\begin{table}[!htb]
\caption{
Power indices: The thin shell wall models are characterized by the index $\alpha$, 
where the wave height is described as $\Phi=\Phi_0 (t/t_0)^{-\alpha}$. 
Energy spectra of the output particles are given by $E^{-p}$. }
Cylindrical model:\\
  \begin{tabular}{|l||c|c|c|} \hline
    & $T/T^{(e)}=0.1$ & $T/T^{(e)}=0.5$ & $T/T^{(e)}=1.0$ \\ 
\hline \hline
    $\alpha=2/3$ & $p=2.5 $& $p=2.5$&$p=2.6$ \\
\hline 
    $\alpha=1/2$ & $p=4.9 $& $p=4.4 $ &$p=4.6 $ \\
 \hline
  \end{tabular}
\vspace{1cm}\\
Spherical model:\\
  \begin{tabular}{|l||c|c|c|} \hline
    & $T/T^{(e)}=0.01$ & $T/T^{(e)}=0.1$ & $T/T^{(e)}=1.0$ \\ 
\hline \hline
    $\alpha=4/3$ & $p=0.84$ & $p=0.84$ & $p=0.87$ \\
\hline 
    $\alpha=1$ & $p=2.1 $& $p=2.0$ &$p=2.1$ \\
 \hline
  \end{tabular}
\label{table:cylindrical}
\end{table}

To clarify which part of energy in initial particle distribution 
contributes the output energy spectrum, 
we divide initial particles into three groups: 

\begin{align}
	\mbox{(i)}&~E \leq \Phi_0/2~&&(v_0 \leq\sqrt{\Phi_0/M}), \\
	\mbox{(ii)}&~ \Phi_0/2 < E \leq \Phi_0 ~&&(\sqrt{\Phi_0/M}< v_0 \leq \sqrt{2\Phi_0/M}),\\
	\mbox{(iii)}&~ \Phi_0<E ~&&(\sqrt{2\Phi_0/M}<v_0 ),
\end{align}
by the initial kinetic energy
(see Fig. \ref{fig:init_dist}).
In the spherical model with $\alpha=4/3$, we calculate acceleration of particles 
and obtain the output energy spectrum as shown in Fig. \ref{fig:colored_spectrum}.
We can see that the lowest energy group (i) contributes the higher part of 
the output energy spectrum. Since the particles with higher energy than $\Phi_0$ cannot be 
trapped by the thin shell wall, then the particles in higher initial energy group (iii) 
are not accelerated effectively. 
Further, we find that law initial energy groups (i) and (ii) make 
some peaks in lower range of the output energy with interval $\mathit\Delta(v/c_0) \sim 2$. 
The particles gain the velocity by $2c_0$ for each reflection;  
then these peaks correspond to the numbers of reflections of test particles by the shrinking shell 
with the speed $c_0$. 

\begin{figure}[!htbp]
\begin{center}
\includegraphics[height=4.3cm]{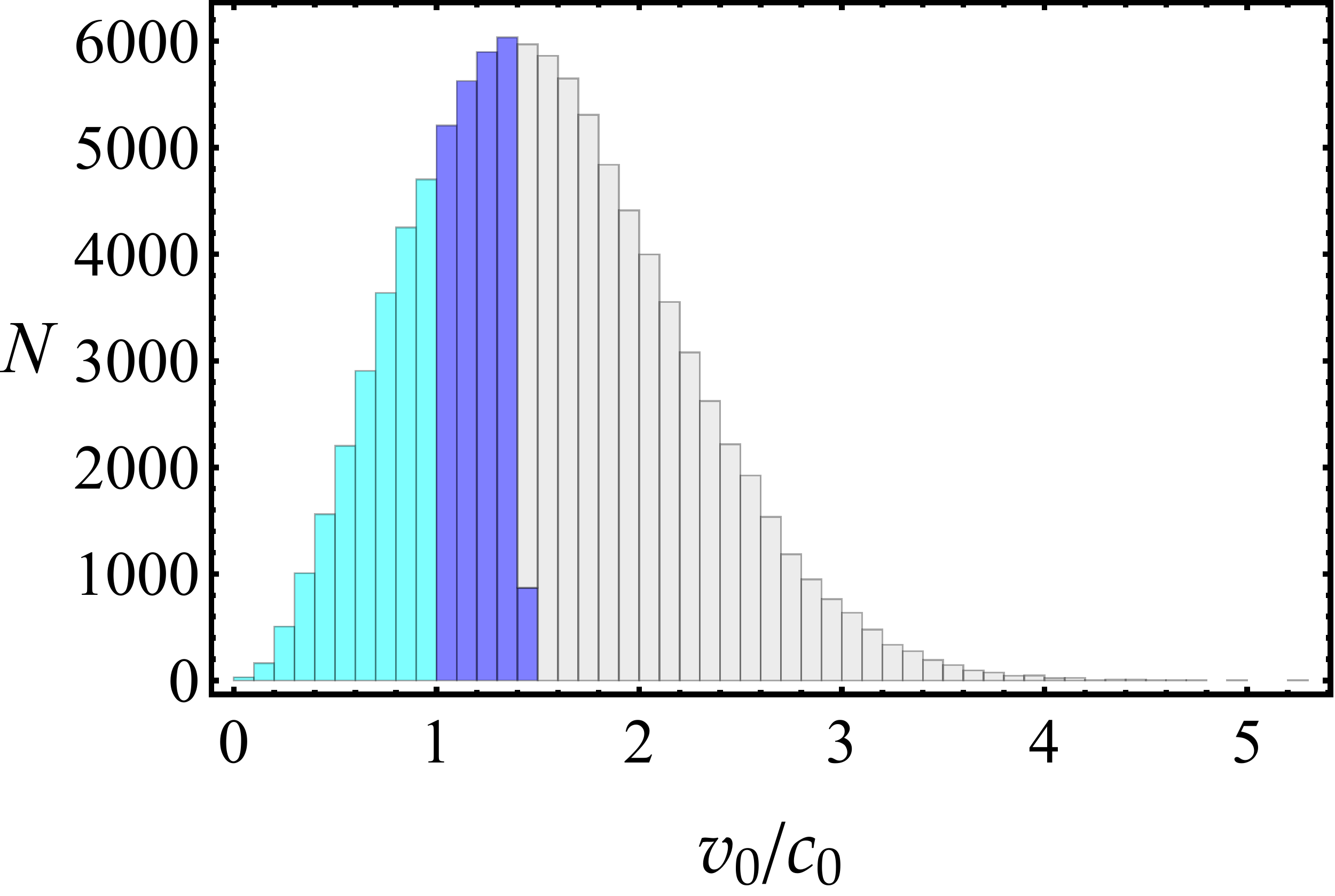}
\caption{Initial particles are classified into three groups by energy: 
({\sl i}) $E \leq \Phi_0/2$ (light blue), 
({\sl ii}) $\Phi_0/2 < E \leq \Phi_0$(dark blue), 
({\sl iii}) $\Phi_0<E $ (gray).
}
\label{fig:init_dist}
\end{center}
\end{figure}
\begin{figure}[!htbp]
\begin{center}
\includegraphics[height=4.3cm]{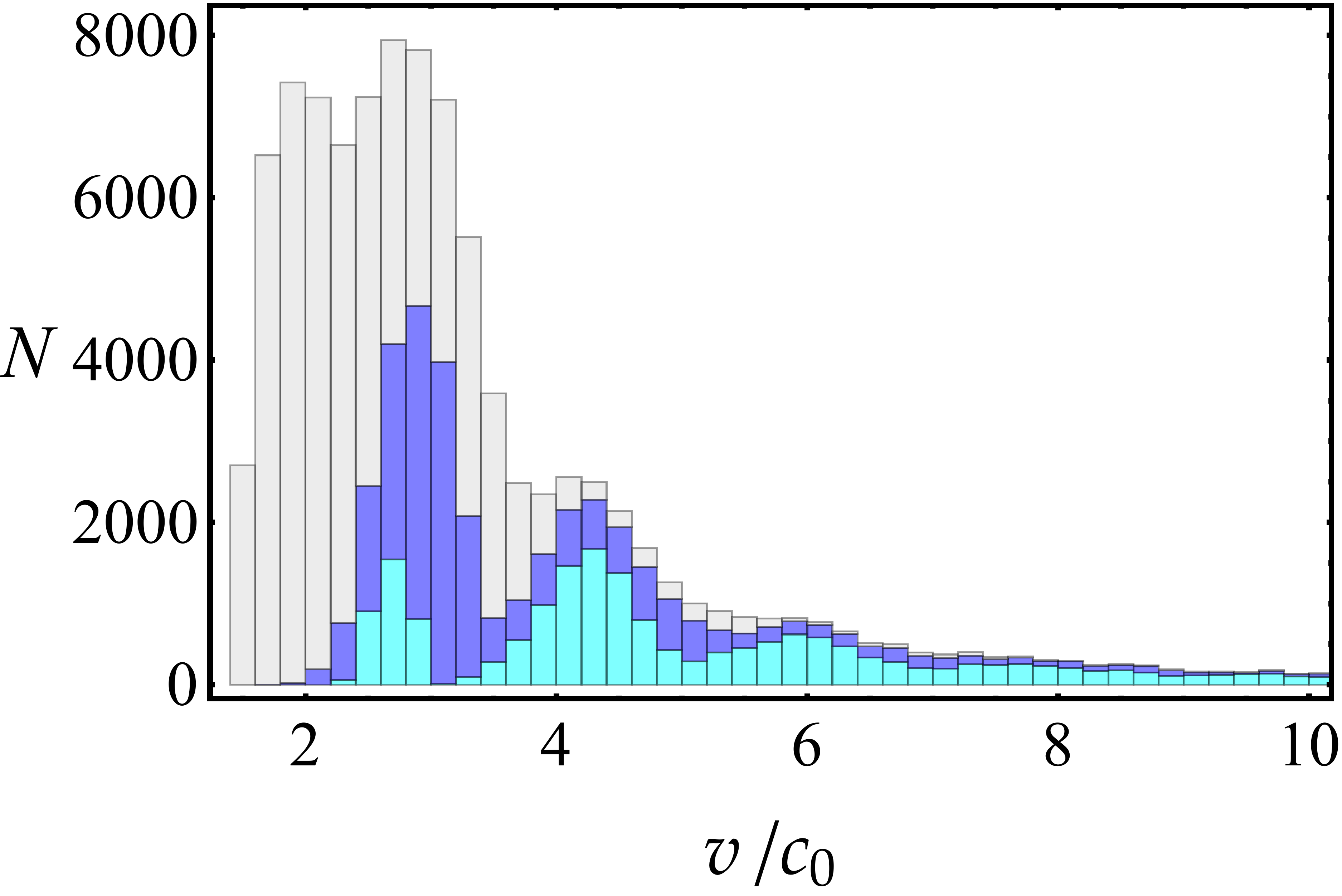}\qquad\qquad
\includegraphics[height=4.3cm]{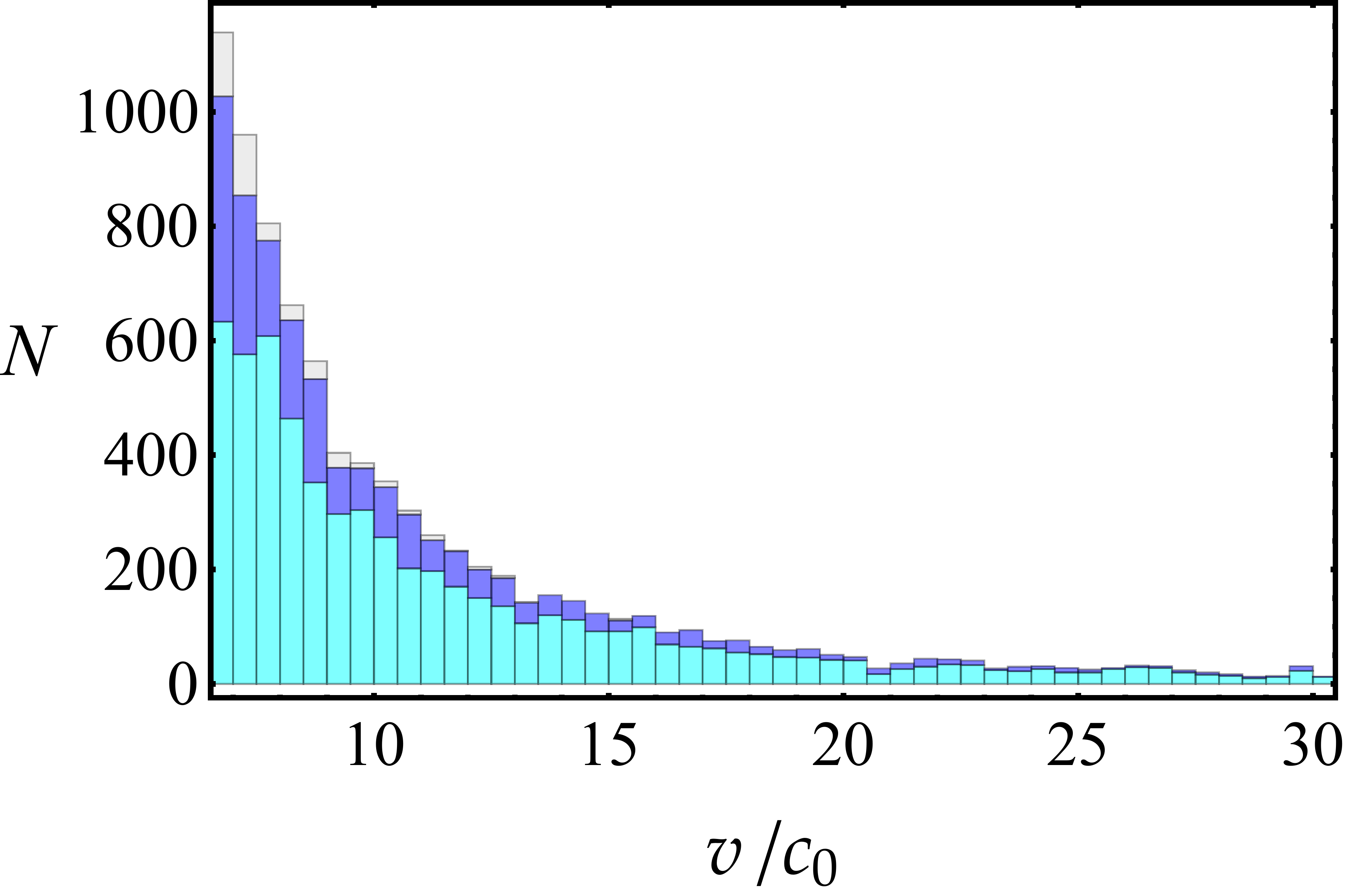}
\caption{Energy distribution of output particles. Particles in groups (i) and (ii) make 
peaks with the interval $\Delta(v_0/c_0) \sim 2$ (left panel). High energy part of distribution 
is shown in the right panel. Almost particles in the high energy part consist of the groups 
(i) and (ii).
}
\label{fig:colored_spectrum}
\end{center}
\end{figure}
%

\section{Summary and Discussion}
\label{summary}

We have investigated a new acceleration mechanism, soliton acceleration, for charged particles 
by using cylindrical or spherical nonlinear acoustic waves propagating in the plasma 
that consists of cold ions and warm electrons. 
We have shown that power law spectra for accelerated output particles are obtained.

The proposed mechanism is different from the Fermi acceleration in the following two points. 
First, in contrast to the Fermi acceleration, where the charged particles are accelerated 
by stochastic reflections by magnetic clouds, in the soliton acceleration, 
the particles are accelerated deterministically 
in a cylindrical or spherical electric potential wall that shrinks with an acoustic soliton. 
In both mechanisms, the power law of energy spectrum of accelerated particles is obtained. 
The reason for the power law in the Fermi acceleration is stochasticity; 
while in the soliton acceleration, the reason is that 
the growth rate of the wave height is the power law in time. 

Secondly, in the Fermi acceleration, only particles with energies that exceed the thermal energy by much 
can cross the shock and can be accelerated. It is not clear what mechanism causes the initial particles 
to have energies sufficiently high. This is the so-called \lq`injection problem''. 
However, particles with the energy less than the initial electric potential energy are accelerated effectively 
in the soliton acceleration. Therefore there is no injection problem in the present mechanism.

We expect that the soliton acceleration mechanism presented in this article can apply 
to the high energy protons of cosmic rays. 
For example, we try to apply to the high energy protons coming from the Sun. 
The high energy protons with the energy range from MeV to GeV 
are observed when the solar flare occurs \cite{Mewaldt}. 
The solar flare is an energetic electromagnetic phenomenon in a short time scale. 
It is widely considered that reconnection of the magnetic field lines occurs 
during solar flare activities \cite{Aschwanden}. 
In the magnetic reconnection region, where the footpoint region of the flare at the chromosphere is,   
the plasma density decreases by coronal mass ejection, and the magnetic field becomes negligibly small. 
The cylindrical or spherical solitons of ion-acoustic waves with the size of reconnection region 
would be excited there \cite{Song-Wu}. Here, we ignore the magnetic field and plasma bulk flow.

We set the temperature of the solar plasma as 
$T^{(e)}_\odot =1 \sim 100 {\rm eV}$,
and the number density of electrons as $n_0=10^{15}\sim 10^{16} {\rm m}^{-3}$, 
then the Debye length as $\lambda_D=10^{-4} \sim 10^{-3}$ m, 
and  the sound velocity $c_0=10^4 \sim 10^5$ m/sec 
for a flare region in the solar atmosphere. 
The radius of the initial wave is assumed to be the size of the reconnection region: 
$r_0 =10^4 {\rm m}=10^7 \sim 10^8 \lambda_D$ \cite{Song-Wu}.
The time scale of the soliton acceleration is given by the initial
radius of the wave divided by the speed of the wave, namely $10^{-1}\sim 1$ sec.
We assumed that the injection energy of the particles is of the same order 
as the thermal energy of the solar atmosphere, i.e., $k_B T^{(e)}_\odot$. 

According to the numerical calculation by the shell models in the previous section,
the output energy spectrum is power law $E^{-p}$ with the index 
$p= 2.5 \sim 4.9$ for the cylindrical model, 
and $p= 0.8 \sim 2.1$ for the spherical model (see Table~\ref{table:cylindrical}). 
If the model is applicable till the cylindrical or spherical shell wall shrinks 
to the size of Debye length, 
the maximum energy is estimated as 
\begin{align}
	E_{\rm max} &\approx \Phi_0 \left(\frac{t_f}{t_0}\right)^{-\alpha}
			= k_B T^{(e)} \left(\frac{r_0}{\lambda_D}\right)^{\alpha} \approx 2~ {\rm GeV}\sim 5 ~{\rm TeV}, 
\label{output_energy}
\end{align}
where the number of input particles is assumed to be large enough. 
Our model would be a candidate for origin of the solar cosmic rays energetically. 

We have found that the growth rate of the wave height of a soliton changes in time 
from the initial stage to the final stage (see Fig.\ref{fig:t_dep_wave_height}). 
If we take the change of growth rate into account, we can consider hybrid shell models, i.e., 
\begin{align}
	&\mbox{\rm Cylindrical shell model,}
\cr
	&\qquad \mbox{initially}: \quad \Phi \propto (t/t_0)^{-2/3};\quad 
\mbox{finally}: \quad\Phi \propto (t/t_0)^{-1/2}, 
\\
	&\mbox{\rm Spherical shell model,}
\cr
	&\qquad\mbox{initially}: \quad \Phi \propto (t/t_0)^{-4/3};\quad 
\mbox{finally}: \quad\Phi \propto (t/t_0)^{-1}, 
\end{align}   
to the solar cosmic rays. 
In the hybrid models, we obtain the energy spectrum shown in 
Fig. \ref{fig:hybrid_spectrum}.
If the double power law reported in Ref.\cite{Mewaldt} should be explained by the acceleration mechanism, 
the soliton acceleration, which leads the double power law naturally, would be a hopeful candidate.

\begin{figure}[!ht]
\begin{center}
\includegraphics[height=5cm]{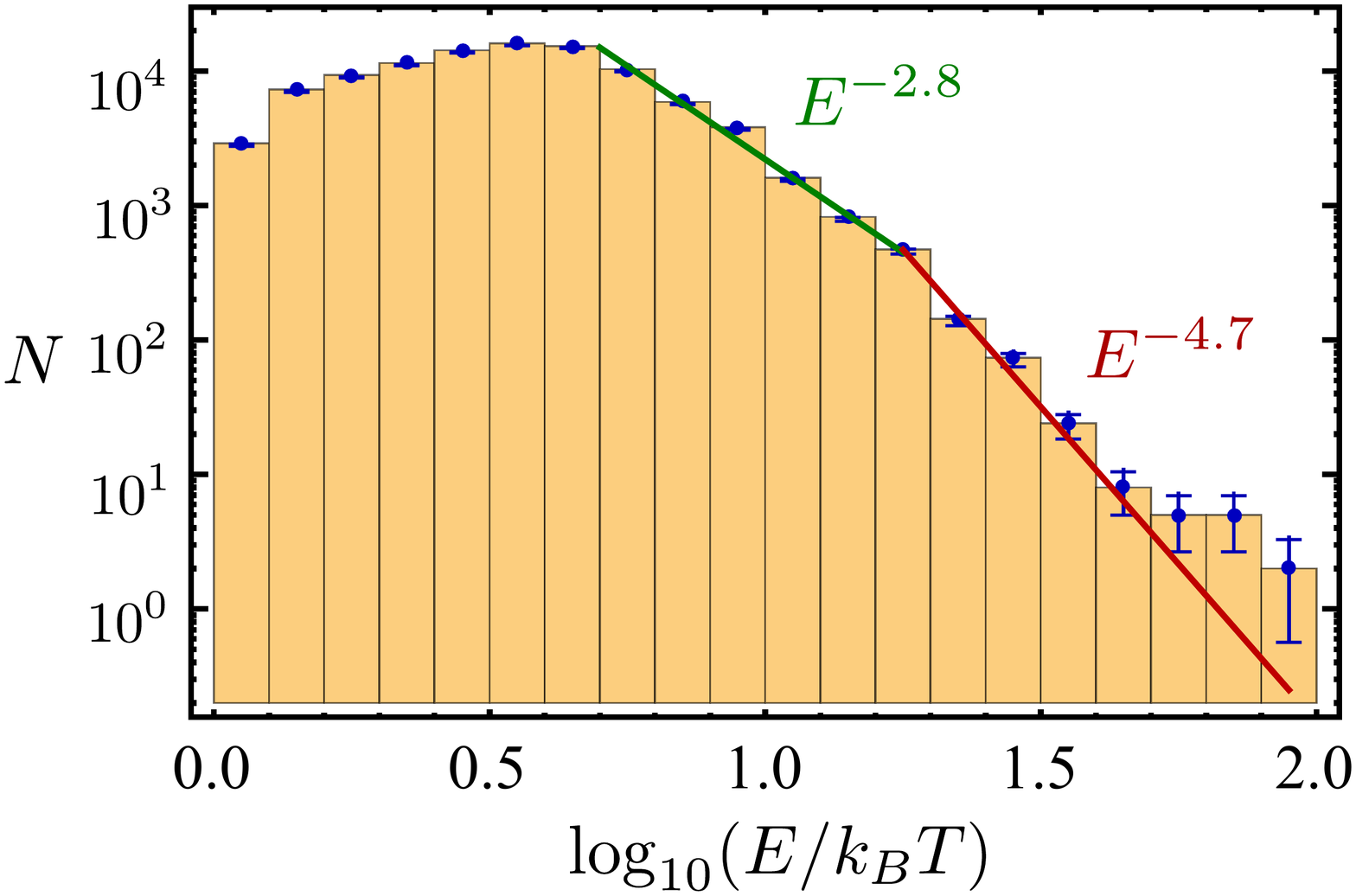} \qquad\qquad
\includegraphics[height=5cm]{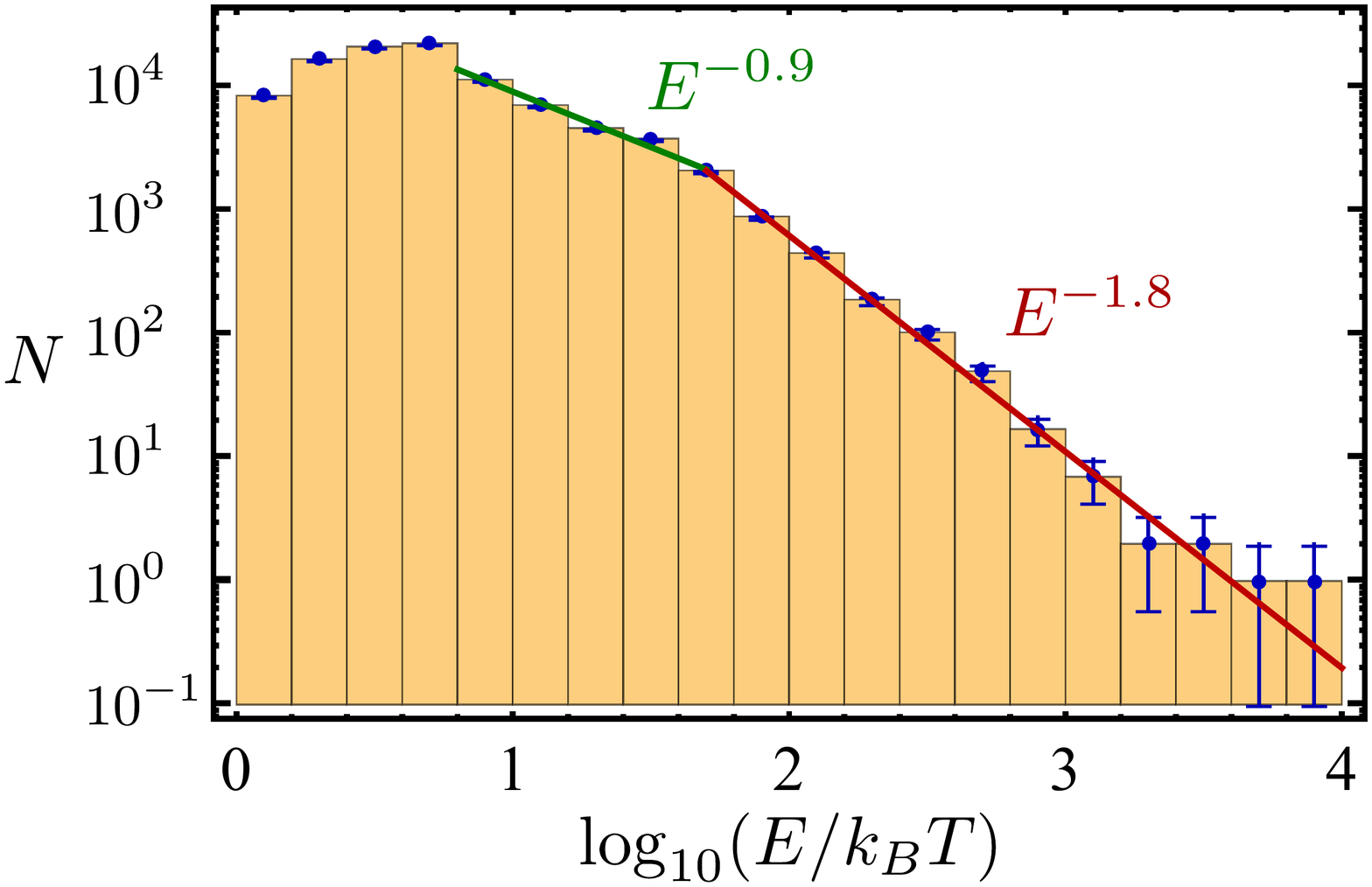}
\caption{Double power spectrum in the cylindrical shell model (left panel). 
The power indices $p=2.8$ in the low energy side and $p=4.7$ in the high energy side.  
The same one in the spherical shell model (right panel). 
The power indices $p=0.9$ in the low energy side and $p=1.8$ in the high energy side.  
}
\label{fig:hybrid_spectrum}
\end{center}
\end{figure}

The maximum value of output energy \eqref{output_energy} exceeds 
the observed value of solar cosmic rays~\cite{Mewaldt}. 
In realistic cases, the final size of the wave would be much larger than the Debye length, and the KdV 
description, which is obtained by the reductive perturbation method for weakly nonlinear waves,  
would break down due to the full nonlinearity of the waves in the final stage. 
We have used the solution of the KdV equation to the highly nonlinear stage in this work 
in order to understand fundamental properties of the particle acceleration mechanism.  
One of the necessary properties for the acceleration mechanism by the waves 
with the electric potential is growth of the amplitude with a power law in time 
as the waves shrink. 
To explain the observed data of the solar cosmic rays, 
it is important to investigate the fully nonlinear solutions for 
the ion-acoustic wave rather than the weakly nonlinear wave solution described by 
the KdV equation~\cite{Webb-Burrows-Ao-Zank}.

The soliton acceleration mechanism presents the source of high energy particles 
at a footpoint of a solar flare. To explain observed data of the solar cosmic rays, 
we should consider the escape process of the high energy particles 
from the solar atmosphere, and the propagation process toward a detector on the earth, and so on. 
These processes would reduce the energy and modify the spectrum of particles produced by the mechanism. 
However, the discussion of these processes is beyond the scope of this article.

In this work, we neglect the magnetic field for simplicity. 
In most astrophysical phenomena the magnetic field plays important roles. 
It would be possible to generalize the soliton acceleration mechanism proposed in this paper 
in the environment of nonvanishing magnetic field.
We will study this issue in the next work.

\section*{Acknowledgment}
The authors would like to thank 
Dr. Ken-ichi Nakao, Dr. Hiromitsu Hamabata, and Dr. Youhei Masada for valuable discussions.  
H.I. was supported by JSPS KAKENHI Grant Number 16K05358. 
M.T. was supported by JSPS KAKENHI Grant Number 17K05439, and DAIKO FOUNDATION.

\end{document}